\documentclass[letterpaper,10pt,conference]{ieeeconf}  
\IEEEoverridecommandlockouts                              
\usepackage{xspace}
\usepackage[english]{babel}
\usepackage[latin1]{inputenc}

\usepackage{amsmath,amssymb}
\usepackage{enumerate}

\usepackage{tikz}
\usetikzlibrary{automata,positioning}
\tikzset{initial text={},
    every state/.style={circle,minimum size=.4cm,draw=blue!50,very thick,fill=blue!20},
    secret/.style={minimum size=.4cm,draw=red!50,very thick,fill=red!20,rectangle},
    node distance=1.5cm,on grid,auto,
    bend angle=65}

\def\ie{{i.e.},~}
\def\eg{{e.g.},~}
\def\st{{s.t.}~}
\def\emptyset{\varnothing}

\def\aonemove{\textit{$A_1$Move}}

\def\Trace{\textit{Tr}}
\def\trace{\textit{tr}}

\def\faulty{\textit{Faulty}}

\def\nonfaulty{\textit{NonFaulty}}

\def\runs{\textit{Runs}} \def\lang{{\cal L}}  
\newcommand{\proj}[1]{\boldsymbol{\pi}_{/#1}} 

\newcommand{\vect}[1]{\mathbf{#1}}

\newtheorem{prob}{Problem}  
\newtheorem{definition}{Definition} 
\newtheorem{theorem}{Theorem} 
\newtheorem{remark}{Remark} 
\newtheorem{example}{Example} 
\newtheorem{corollary}{Corollary}

\newcommand{\setN}{\mathbb N}
\newcommand{\setR}{\mathbb R}
\newcommand{\setB}{\mathbb B}
\newcommand{\setZ}{\mathbb Z}
\newcommand{\setQ}{\mathbb Q}

\def\calA{{\cal A}}
\def\calB{{\cal B}}
\def\calD{{\cal D}}

\def\calC{{\cal C}}

\def\cc{\calC}

\def\last{\textit{last}}

\def\endef{\ifmmode\squareforged\else{\unskip\nobreak\hfil
\penalty50\hskip1em\null\nobreak\hfil$\blacksquare$
\parfillskip=0pt\finalhyphendemerits=0\endgraf}\fi}

\def\ssi{iff\xspace}

\def\aonemove{\textit{$A_1$Move}}

\def\tauac{\tau}

\newcommand{\dur}{{\textit{Dur}}} 
\def\inv{\textit{Inv}}

\def\tw{\textit{TW\/}}

\def\dive{\textit{Div}} 
\def\untimed{\textit{Unt}}

\def\true{\mbox{\textsc{true}}}
\def\false{\mbox{\textsc{false}}}

\def\rg{\textit{RG}}

\title{\LARGE \bf A Note on  Fault Diagnosis Algorithms}

\author{Franck Cassez, \IEEEmembership{Member, IEEE}%
  \thanks{Franck Cassez is with National ICT Australia \& CNRS, Locked Bag 6016, The
    University of New South Wales, Sydney NSW~1466, Australia. \texttt{\scriptsize franck.cassez@cnrs.irccyn.fr,Franck.Cassez@nicta.com.au}}%
  \thanks{Author supported by a Marie Curie International
    Outgoing Fellowship within the 7th
    European Community Framework Programme.} 
}

\begin{document}
\maketitle
  
\thispagestyle{empty}

\begin{abstract} 
  In this paper we review algorithms for checking diagnosability of
  discrete-event systems and timed automata. We point out that the
  diagnosability problems in both cases reduce to the emptiness
  problem for (timed) B\"uchi automata. Moreover, it is known that,
  checking whether a discrete-event system is diagnosable, can also be
  reduced to checking bounded diagnosability. We establish a similar
  result for timed automata.  We also provide a synthesis of the
  complexity results for the different fault diagnosis problems.
\end{abstract}
\medskip

\textbf{Note:} This paper is an extended version of the paper
pub\-lis\-hed in the proceedings of CDC'09.

\section{Introduction}

Discrete-event systems~\cite{RW87,RW89} (DES) can be modelled by
finite automata over an alphabet of \emph{observable} events $\Sigma$.
To address decision problems under \emph{partial observation} of DES,
it is sufficient to add a special event $\tauac$ which represents all
the \emph{unobservable} actions.

The \emph{Fault diagnosis problem} is a typical example of
a problem under partial observation. 
We assume that the behavior of the DES is known and a model of it is
available as a finite automaton over an alphabet $\Sigma \cup
\{\tauac,f\}$, where $\Sigma$ is the set of observable events,
$\tauac$ represents the unobservable events, and $f$ is a special
unobservable event that corresponds to the faults: this is the
original framework introduced by
M.~Sampath~and~\emph{al.}~\cite{Raja95} and the reader is referred to
this paper for a clear and exhaustive introduction to the
subject\footnote{The ``companion paper''~\cite{Raja96} focuses on a
  less stringent notion of diagnosability called I-diagnosability.}.
The aim of fault diagnosis is to detect \emph{faulty} sequences of the
DES by observing only the events in $\Sigma$.  A \emph{faulty}
sequence is a sequence of the DES containing an occurrence of event
$f$.  We assume that an \emph{observer} which has to detect faults,
knows the specification/model of the DES, and it is able to observe
sequences of \emph{observable} events.  Based on this knowledge, it
has to announce whether an observation (a word in $\Sigma^*$) was
produced by a faulty sequence (in $(\Sigma \cup \{\tauac,f\})^*$) 
or not.
A \emph{diagnoser} (for a DES) is an observer which observes the
sequences of observable events and is able to detect whether a fault
event occurred, although it is not observable.  If a diagnoser can
detect a fault at most $\Delta$ steps\footnote{Steps are measured by
  the number of events, observable or not, which occur in the DES.}
after it occurred, the DES is said to be $\Delta$-diagnosable. It is
diagnosable if it is $\Delta$-diagnosable for some $\Delta \in \setN$.
Checking whether a DES is $\Delta$-diagnosable for a given $\Delta$ is
called the \emph{bounded diagnosability problem}; checking whether a
DES is diagnosable is the \emph{diagnosability problem}.

Checking {\em diagnosability} for a given DES and a fixed set of
observable events can be done in polynomial time using the algorithms
of~\cite{Jiang-01,yoo-lafortune-tac-02}.  Nevertheless the size of the
diagnoser can be exponential as it involves a determinization step.
The extension of this DES framework to timed
automata~\cite{AlurDill94} (TA) has been proposed by
S.~Tripakis~\cite{tripakis-02}, and he proved that the problem of
checking diagnosability of a timed automaton is PSPACE-complete.  In
the timed case, the diagnoser may be a Turing machine.  In a
subsequent work by P.~Bouyer~and~\emph{al.}~\cite{Bouyer-05}, the
problem of checking whether a timed automaton is diagnosable by a
diagnoser which is a \emph{deterministic} timed automaton was studied
(we will not refer to this work in this paper.)

The algorithms proposed in the DES
framework~\cite{Jiang-01,yoo-lafortune-tac-02} and in the timed
automata framework~\cite{tripakis-02} rely on different assumptions
and use different techniques: for
example~\cite{Jiang-01,yoo-lafortune-tac-02} assumes that the DES is
\emph{live} and contains no unobservable loops; the algorithm to check
the diagnosability problem then consists in checking whether a cycle
exists in a suitable product automaton; the algorithm
of~\cite{tripakis-02} for timed automata consists in checking whether
a infinite word can be accepted by a (product) B\"uchi automaton: the
main reason for the use of a B\"uchi acceptance condition in this case
is to ensure time divergence.

\noindent{\it \bfseries Our Contribution.} 
In this paper, we try to put into perspective the results
of~\cite{Jiang-01,yoo-lafortune-tac-02,tripakis-02,cassez-fi-08} by
giving a uniform presentation of the algorithms for fault diagnosis
both in the DES and timed automata settings.  We also establish a (not
difficult but still) missing result for timed automata: diagnosability
can be reduced to bounded diagnosability. Another contribution of this
paper is to examine in details the complexity of the problems and this
is summarized in Table~\ref{tab-summary}.

The results in this paper that are not new and have already been
published are followed by the reference(s) in the core of the text of
after the Theorem keyword.

One such result is Theorem~\ref{thm-diagnosability} which already
appeared in~\cite{cassez-fi-08}.  It generalizes the previous results
of~\cite{Jiang-01,yoo-lafortune-tac-02} and shows that fault diagnosis
reduces to B\"uchi emptiness for DES.  This has some interesting
consequences regarding the algorithmic aspects of the problem as well
as the tools that can be used to verify diagnosability. These
considerations (Section~\ref{sec-des}) might be of interest for the
DES community.

\noindent{\it \bfseries Organisation of the Paper.} 
Section~\ref{sec-prelim} recalls the definitions of timed automata.
Section~\ref{sec-fd} introduces the fault diagnosis problems we are
interested in.  Sections~\ref{sec-des} and~\ref{sec-ta} describes the
algorithms to solve the diagnosability problems respectively for DES
and TA.  Section~\ref{sec-conclu} summarizes the results.


\section{Preliminaries}\label{sec-prelim}
$\Sigma$ denotes a finite alphabet and $\Sigma_\tauac=\Sigma \cup
\{\tauac\}$ where $\tauac \not\in \Sigma$ is the \emph{unobservable}
action.  $\setB=\{\true,\false\}$ is the set of boolean values,
$\setN$ the set of natural numbers, $\setZ$ the set of integers and
$\setQ$ the set of rational numbers.  $\setR$ is the set of real
numbers and $\setR_{\geq 0}$ is the non-negative real numbers.

\subsection{Clock Constraints}
Let $X$ be a finite set of variables called \emph{clocks}.  A
\emph{clock valuation} is a mapping $v : X \rightarrow \setR_{\geq
  0}$. We let $\setR_{\geq 0}^X$ be the set of clock valuations over
$X$. We let $\vect{0}_X$ be the \emph{zero} valuation where all the
clocks in $X$ are set to $0$ (we use $\vect{0}$ when $X$ is clear from
the context).  Given $\delta \in \setR$, $v + \delta$ denotes the
valuation defined by $(v + \delta)(x)=v(x) + \delta$. We let $\cc(X)$
be the set of \emph{convex constraints} on $X$, \ie the set of
conjunctions of constraints of the form $x \bowtie c$ with $c
\in\setZ$ and $\bowtie \in \{\leq,<,=,>,\geq\}$. Given a constraint $g
\in \cc(X)$ and a valuation $v$, we write $v \models g$ if $g$ is
satisfied by $v$.  Given $R \subseteq X$ and a valuation $v$, $v[R]$
is the valuation defined by $v[R](x)=v(x)$ if $x \not\in R$ and
$v[R](x)=0$ otherwise.

\subsection{Timed Words}
The set of finite (resp. infinite) words over $\Sigma$ is $\Sigma^*$
(resp. $\Sigma^\omega$) and we let $\Sigma^\infty=\Sigma^* \cup \Sigma
^\omega$. A \emph{language} $L$ is any subset of $\Sigma^\infty$. A
finite (resp. infinite) \emph{timed word} over $\Sigma$ is a word in
$(\setR_{\geq 0}.\Sigma)^*.\setR_{\geq 0}$ (resp. $(\setR_{\geq
  0}.\Sigma)^\omega$).  We let $\dur(w)$ be the duration of a timed
word $w$ which is defined to be the sum of the durations (in
$\setR_{\geq 0}$) which appear in $w$; if this sum is infinite, the
duration is $\infty$.  Note that the duration of an infinite word can
be finite, and such words which contain an infinite number of letters,
are called \emph{Zeno} words.  We let $\untimed(w)$ be the
\emph{untimed} version of $w$ obtained by erasing all the durations in
$w$, \eg $\untimed(0.4\ a\ 1.0\ b\ 2.7 \ c)=abc$.
%
%
In this paper we write timed words as $0.4\ a\ 1.0\ b\ 2.7 \ c \cdots$
where the real values are the durations elapsed between two letters:
thus $c$ occurs at global time $4.1$. 

$\tw^*(\Sigma)$ is the set of finite timed words over $\Sigma$,
$\tw^\omega(\Sigma)$, the set of infinite timed words and
$\tw^\infty(\Sigma)=\tw^*(\Sigma) \cup \tw^\omega(\Sigma)$. A
\emph{timed language} is any subset of $\tw^\infty(\Sigma)$.

Let $\proj{\Sigma'}$ be the projection of timed words of
$\tw^\infty(\Sigma)$ over timed words of $\tw^\infty(\Sigma')$.  When
projecting a timed word $w$ on a sub-alphabet $\Sigma' \subseteq
\Sigma$, the durations elap\-sed bet\-ween two events are set
accordingly: for instance $\proj{\{a,c\}}(0.4 \ a\ 1.0\ b\ 2.7 \ c
)=0.4 \ a \ 3.7 \ c$ (projection erases some letters but
keep the time elapsed between two letters).  Given a timed language
$L$, we let $\untimed(L)=\{ \untimed(w) \ | \ w \in L \}$.  Given
$\Sigma' \subseteq \Sigma$, $\proj{\Sigma'}(L)=\{ \proj{\Sigma'}(w) \
| \ w \in L\}$.

\subsection{Timed Automata}
Timed automata (TA) are finite automata extended with real-valued clocks to
specify timing constraints between occurrences of events.  For a
detailed presentation of the fundamental results for timed automata,
the reader is referred to the seminal paper of R.~Alur and
D.~Dill~\cite{AlurDill94}.
\noindent\begin{definition}[Timed Automaton]\label{def-ta} 
  A \emph{Timed Automaton} $A$ is a tuple $(L,$ $l_0,$
  $X,\Sigma_\tauac, E, \inv, F, R)$ where:
$L$ is a finite set of  \emph{locations}; 
$l_0$ is the \emph{initial location};
$X$ is a finite set of \emph{clocks};
$\Sigma$ is a finite set of \emph{actions}; 
$E \subseteq L \times\calC(X) \times \Sigma_\tauac \times 2^X \times
L$ is a finite set of \emph{transitions}; for
$(\ell,g,a,r,\ell') \in E$, $g$ is the \emph{guard}, $a$ the \emph{action},
and $r$ the \emph{reset} set;
$\inv \in \calC(X)^L$ associates with each location an
  \emph{invariant}; as usual we require the invariants to be
  conjunctions of constraints of the form $x \preceq c$ with $\preceq \in
  \{<,\leq\}$.
  $F \subseteq L$ and $R \subseteq L$ are respectively the
  \emph{final} and \emph{repeated} sets of locations. \endef
\end{definition}
An example of TA is given in Fig.~\ref{fig-ex-diag1}.  A
\emph{state} of $A$ is a pair $(\ell,v) \in L \times \setR_{\geq
  0}^X$.
%
%
A \emph{run} $\varrho$ of $A$ from $(\ell_0,v_0)$ is a (finite or
infinite) sequence of alternating \emph{delay} and \emph{discrete}
moves:
\begin{eqnarray*}
  \varrho & = & (\ell_0,v_0) \xrightarrow{\delta_0} (\ell_0,v_0 + \delta_0)
  \xrightarrow{a_0} (\ell_1,v_1) \cdots  \\ & & \cdots \xrightarrow{a_{n-1}} (\ell_n,v_n)
  \xrightarrow{\delta_n} (\ell_n,v_n+ \delta_n) \cdots 
\end{eqnarray*}
\st for every $i \geq 0$:
\begin{itemize}
\item $v_i + \delta \models \inv(\ell_i)$ for $0 \leq \delta \leq \delta_i$;
\item there is some transition $(\ell_i,g_i,a_i,r_i,\ell_{i+1}) \in E$
  \st: ($i$) $v_i + \delta_i \models g_i$ and ($ii$)
  $v_{i+1}=(v_i+\delta_i)[r_i]$.
\end{itemize}
The set of finite (resp. infinite) runs 
from a state $s$ is
denoted $\runs^*(s,A)$ (resp. $\runs^\omega(s,A)$) and we define
$\runs^*(A)=\runs^*((l_0,\vect{0}),A)$ and
$\runs^\omega(A)=\runs^\omega((l_0,\vect{0}),A)$.  As before 
$\runs(A)=\runs^*(A) \cup \runs^\omega(A)$.  If $\varrho$ is finite
and ends in $s_n$, we let $\last(\varrho)=s_n$.  Because of the
den\-se\-ness of the time domain, the unfolding of $A$ as a graph is
infinite (uncountable number of states and delay edges).
The \emph{trace}, $\trace(\varrho)$, of a run $\varrho$ is the timed
word $\proj{\Sigma}(\delta_0 a_0 \delta_1 a_1 \cdots a_n \delta_n
\cdots)$.  We let
$\dur(\varrho)=\dur(\trace(\varrho))$.  For $V \subseteq \runs(A)$, we
let $\Trace(V)=\{\trace(\varrho) \ | \ \textit{ $\varrho \in V$}\}$,
which is the set of traces of the runs in $V$.

A finite (resp. infinite) timed word $w$ is \emph{accepted} by $A$ if
it is the trace of a run of $A$ that ends in an $F$-location (resp. a
run that reaches infinitely often an $R$-location).  $\lang^*(A)$
(resp. $\lang^\omega(A)$) is the set of traces of finite
(resp. infinite) timed words accepted by $A$, and $\lang(A)=\lang^*(A)
\cup \lang^\omega(A)$ is the set of timed words accepted by $A$.
In the sequel we often omit the sets $R$ and $F$ in TA and this
implicitly means $F=L$ and $R=\emptyset$.

\medskip

A finite automaton (FA) is a particular TA with $X=\emptyset$.
Consequently guards and invariants are vacuously true and time
elapsing transitions do not exist.  We write $A=(L,$
$l_0,\Sigma_\tauac,E,F,R)$ for a FA.  A run is thus a
sequence of the form:
\begin{eqnarray*}
  \varrho & = & \ell_0 
  \xrightarrow{a_0} \ell_1 \cdots   \cdots \xrightarrow{a_{n-1}} \ell_n
  \cdots 
\end{eqnarray*}
where for each $i \geq 0$, $(\ell_i,a_i,\ell_{i+1}) \in E$.
Definitions of traces and languages are straightforward.  In this
case, the duration of a run $\varrho$ is the number of steps
(including $\tauac$-steps) of $\varrho$: if $\varrho$ is finite and
ends in $\ell_n$, $\dur(\varrho)=n$ and otherwise
$\dur(\varrho)=\infty$.

\subsection{Region Graph of a TA}
The \emph{region graph} $\rg(A)$ of a TA $A$ is a finite quotient of
the infinite graph of $A$ which is time-abstract bisimilar to
$A$~\cite{AlurDill94}.  It is a FA on the alphabet $E'= E \cup
\{\tauac\}$. The states of $\rg(A)$ are pairs $(\ell,r)$ where $\ell
\in L$ is a location of $A$ and $r$ is a \emph{region} of $\setR_{\geq
  0}^X$. More generally, the edges of the graph are tuples $(s,t,s')$
where $s,s'$ are states of $\rg(A)$ and $t \in E'$.  Genuine
unobservable moves of $A$ labelled $\tauac$ are labelled by tuples of
the form $(s,(g,\tauac,r),s')$ in $\rg(A)$.
An edge $(g,\lambda,R)$ in the region graph corresponds to a discrete
transition of $A$ with guard $g$, action $\lambda$ and reset set $R$.
A $\tauac$ move in $\rg(A)$ stands for a delay move to the
time-successor region.  The initial state of $\rg(A)$ is
$(l_0,\vect{0})$.  A final (resp. repeated) state of $\rg(A)$ is a
state $(\ell,r)$ with $\ell \in F$ (resp. $\ell \in R$).  A
fundamental property of the region graph~\cite{AlurDill94} is:
\begin{theorem}[\cite{AlurDill94}] \label{thm-alur}
  $\lang(\rg(A))=\untimed(\lang(A))$. 
\end{theorem}
In other words:
  \begin{enumerate}
  \item if $w$ is accepted by $\rg(A)$, then there is a timed word $v$
    with $\untimed(v)=w$ \st $v$ is accepted by $A$.
  \item if $v$ is accepted by $A$, then $\untimed(w)$ is accepted
    $\rg(A)$.
  \end{enumerate}
  The (maximum) size of the region graph is exponential in the number
  of clocks and in the maximum constant of the automaton $A$
  (see~\cite{AlurDill94}): $|\rg(A)|=|L|\cdot |X|! \cdot 2^{|X|} \cdot
  K^{|X|}$ where $K$ is the largest constant used in $A$.

\subsection{Product of TA}
\begin{definition}[Product of TA] \label{def-prod-sync} Let
  $A_i=(L_i,l_0^i,X_i,\Sigma^i_{\tauac},$ $E_i,\inv_i)$, $i \in\{1,2\}$,
  be TA \st $X_1 \cap X_2 = \emptyset$.  The \emph{product} of
  $A_1$ and $A_2$ is the TA $A_1 \times
  A_2=(L,l_0,X,$$\Sigma_{\tauac},E,\inv)$ given by:
$L=L_1 \times L_2$;
$l_0=(l_0^1,l_0^2)$;
$\Sigma=\Sigma^1 \cup \Sigma^2$;
$X = X_1 \cup X_2$; and
$E \subseteq L \times \calC(X) \times \Sigma_\tauac \times 2^X \times
    L$ and
    $((\ell_1,\ell_2),g_{1,2},\sigma,r,(\ell'_1,\ell'_2)) \in E$
    if:
    \begin{itemize}
    \item either $\sigma \in (\Sigma_1 \cap \Sigma_2) \setminus
      \{\tauac \}$, and ($i$) $(\ell_k,g_k,\sigma,r_k,\ell'_k) \in
      E_k$ for $k=1$ and $k=2$; ($ii$) $g_{1,2} = g_1 \wedge g_2$ and
      ($iii$) $r=r_1 \cup r_2$;
    \item or for $k=1$ or $k=2$, $\sigma \in (\Sigma_k \setminus
      \Sigma_{3-k}) \cup \{\tauac\}$, and ($i$)
      $(\ell_k,g_k,\sigma,r_k,\ell'_k) \in E_k$; ($ii$) $g_{1,2}=g_k$
      and ($iii$) $r=r_k$;
    \end{itemize}
    and finally $\inv(\ell_1,\ell_2)= \inv(\ell_1) \wedge
    \inv(\ell_2)$.
     \endef
\end{definition}
The definition of product also applies to finite automata.


\section{Fault Diagnosis Problems}\label{sec-fd}
The material in this section is based
on~\cite{yoo-lafortune-tac-02,tripakis-02,cassez-fi-08}.  To model
timed systems with faults, we use timed automata on the alphabet
$\Sigma_{\tauac,f}=\Sigma_{\tauac}\cup \{f\}$ where $f$ is the
\emph{faulty} (unobservable) event. We only consider one type of fault
here, but the results we give are valid for many types of faults
$\{f_1,f_2, \cdots,f_n\}$: indeed solving the many types
diagnosability problem amounts to solving $n$ one type diagnosability
problems~\cite{yoo-lafortune-tac-02}.
Other unobservable events are abstracted as a $\tauac$ action (one 
$\tauac$ suffices as these events are all unobservable).

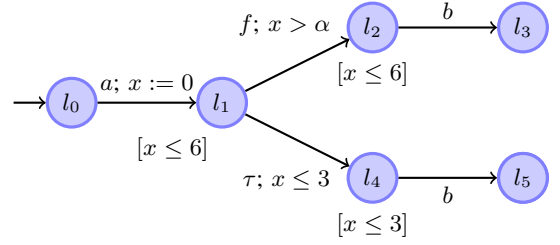
\begin{figure}[hbtp]
  \centering
  \begin{tikzpicture}[thick,node distance=1cm and 2cm]%
    \small
    \node[state,initial] (q_00) {$l_0$}; 
    \node[state] (q_0) [right=of q_00,label=-100:{$[x \leq 6]$}] {$l_1$}; 
    \node[state] (q_1) [above right=of q_0,label=-90:{$[x \leq 6]$}] {$l_2$};
    \node[state] (q_2) [below right=of q_0,label=-90:{$[x \leq 3]$}] {$l_4$};
    \node[state] (q_3) [right=of q_2] {$l_4$};
    \node[state] (q_4) [right=of q_1] {$l_3$};
    \node[state] (q_5) [right=of q_2] {$l_5$};
    \path[->] (q_00) edge node {$a$; $x:=0$} (q_0) 
              (q_0)  edge node[pos=0.9]  {$f$; $x> \alpha$} (q_1)
                    edge [swap,pos=0.9] node  {$\tau$; $x \leq 3$} (q_2)
              (q_1) edge  node  {$b$} (q_4)
              (q_2) edge  [swap] node  {$b$} (q_5);
  \end{tikzpicture}
\caption{The Timed Automaton $\calA(\alpha)$}
\label{fig-ex-diag1}
\end{figure}

The system we want to supervise is given as a TA
$A=(L,l_0,$$X,\Sigma_{\tauac,f},E,\inv)$. Fig.~\ref{fig-ex-diag1}
gives an example of such a system ($\alpha \in \setN$ is a 
parameter). Invariants in the automaton $\calA(\alpha)$ are written
within square brackets as in $[x \leq 3]$.
 
\noindent Let $\Delta \in \setN$. A run of $A$
\begin{eqnarray*}
  \varrho & = & (\ell_0,v_0) \xrightarrow{\delta_0} (\ell_0,v_0 + \delta_0)
  \xrightarrow{a_0} (\ell_1,v_1) \cdots \\ 
  & & \cdots \xrightarrow{a_{n-1}} (\ell_n,v_n)
  \xrightarrow{\delta_n} (\ell_n,v_n+ \delta) \cdots
\end{eqnarray*}
is $\Delta$-faulty if: (1) there is an index $i$ \st $a_i=f$ and (2)
the duration of the run $\varrho'=(\ell_{i},v_i)
\xrightarrow{\delta_{i}} \cdots \xrightarrow{\delta_n}
(\ell_n,v_n+\delta_n) \cdots$ is larger than $\Delta$.  We let
$\faulty_{\geq \Delta}(A)$ be the set of $\Delta$-faulty runs of $A$.
Note that by definition, if $\Delta' \geq \Delta$ then $\faulty_{\geq
  \Delta'}(A) \subseteq \faulty_{\geq \Delta}(A)$. We let
$\faulty(A)=\cup_{\Delta \geq 0}\faulty_{\geq \Delta}(A)=\faulty_{\geq
  0}(A)$ be the set of faulty runs of $A$, and $\nonfaulty(A)=\runs(A)
\setminus \faulty(A)$ be the set of non-faulty runs of $A$.
Finally  
$$\faulty^{\textit{tr}}_{\geq
  \Delta}(A)=\Trace(\faulty_{\geq \Delta}(A))$$ and
$$\nonfaulty^{\textit{tr}}(A)=\Trace(\nonfaulty(A))$$ 
which are the traces\footnote{Notice that $\trace(\varrho)$ erases
  $\tauac$ and $f$.} of $\Delta$-faulty and non-faulty runs of $A$.

The purpose of fault diagnosis is to detect a fault as soon as
possible.  Faults are unobservable and only the events in $\Sigma$ can
be observed as well as the time elapsed between these events.
Whenever the system generates a timed word $w$, the observer can only
see $\proj{\Sigma}(w)$.  If an observer can detect faults in this way
it is called a \emph{diagnoser}.  A diagnoser must detect a fault
within a given delay $\Delta \in \setN$.

\begin{definition}[$\Delta$-Diagnoser]\label{def-diag}
  Let $A$ be a TA over the alphabet $\Sigma_{\tauac,f}$
  and $\Delta \in \setN$.  A \emph{$\Delta$-diagnoser}
  for $A$ is a mapping  $D: \tw^*(\Sigma)\rightarrow \{0,1\}$
  such that:
  \begin{itemize}
  \item for each $\varrho \in \nonfaulty(A)$,
    $D(\trace(\varrho))=0$,
  \item for each $\varrho \in \faulty_{\geq \Delta}(A)$,
    $D(\trace(\varrho))=1$. \endef
  \end{itemize}
\end{definition}
$A$ is $\Delta$-diagnosable if there exists a $\Delta$-diagnoser for
$A$. $A$ is diagnosable if there is some $\Delta \in \setN$ \st $A$
est $\Delta$-diagnosable.

\begin{remark}
  Nothing is required for the $\Delta'$-faulty words with $\Delta' <
  \Delta$. Thus a diagnoser could change its mind and answers $1$ for
  a $\Delta'$-faulty word, and $0$ for a $\Delta''$-faulty word with
  $\Delta' < \Delta'' < \Delta$.
\end{remark}
\begin{example}
  The TA $\calA(3)$ in Fig.~\ref{fig-ex-diag1} taken
  from~\cite{tripakis-02} is $3$-diagnosable.  For the timed words of
  the form $t . a . \delta . b . t'$ with $\delta \leq 3$, no fault
  has occurred, whereas when $\delta > 3$ a fault must have occurred.
  A diagnoser can then be easily constructed. As we have to wait for a
  ``$b$'' action to detect a fault, $D$ cannot detect a fault in $2$
  time units.
  If $\alpha=2$, in $\calA(2)$ there are two runs:
  \begin{eqnarray*}
    \rho_1(\delta) & = & (l_0,0) \xrightarrow{a} (l_1,0) 
    \xrightarrow{2.5} (l_1,2.5) \xrightarrow{f} (l_2,2.5)  \\
    & & \hspace*{0cm} \xrightarrow{0.2} (l_2,2.7) \xrightarrow{b} (l_3,2.7) \xrightarrow{\delta} (l_2,2.7+\delta) \\
    \rho_2(\delta) & = & (l_0,0) \xrightarrow{a} (l_1,0) 
    \xrightarrow{2.5} (l_1,2.5) \xrightarrow{\tau} (l_4,2.5)  \\
    & & \hspace*{0cm} \xrightarrow{0.2} (l_4,2.7) \xrightarrow{b} (l_5,2.7) \xrightarrow{\delta} (l_5,2.7+\delta)\\
  \end{eqnarray*}
  that satisfy
  $\trace(\rho_1(\delta))=\trace(\rho_2(\delta))$,
  and this for every $\delta \geq 0$. For each $\Delta \in \setN$,
  there are two runs $\rho_1(\Delta)$ and $\rho_2(\Delta)$ which
  produce the same observations and thus no diagnoser can
  exist. $\calA(2)$ is not diagnosable.
\end{example}

\noindent The classical fault diagnosis problems are the following:
\smallskip
\begin{prob}[Bounded or $\Delta$-Diagnosability] \label{prob-delta-diag} \mbox{} \\
  \textsc{Inputs:} A  TA $A=(L,\ell_0,X,\Sigma_{\tauac,f},E,\inv)$ and $\Delta \in \setN$. \\
  \textsc{Problem:} Is $A$ $\Delta$-diagnosable?
\end{prob}
\begin{prob}[Diagnosability] \label{prob-diag} \mbox{} \\
  \textsc{Inputs:} A TA $A=(L,\ell_0,X,\Sigma_{\tauac,f},E,\inv)$. \\
  \textsc{Problem:} Is $A$ diagnosable?
\end{prob}
\begin{prob}[Maximum delay] \label{prob-delay} \mbox{} \\
  \textsc{Inputs:} A TA $A=(L,\ell_0,X,\Sigma_{\tauac,f},E,\inv)$. \\
  \textsc{Problem:} If $A$ is diagnosable, what is the minimum
  $\Delta$ \st $A$ is $\Delta$-diagnosable ?
\end{prob}
\smallskip

We do not address here the problem of synthesizing a diagnoser and the
reader is referred
to~\cite{yoo-lafortune-tac-02,Jiang-01,tripakis-02,Bouyer-05} for a
detailed presentation.

A necessary and sufficient condition for diagnosability was already
established in~\cite{Raja95}, but was stated on a candidate diagnoser.
We give here a simple language based condition, valid in both the
discrete and timed cases.  According to Definition~\ref{def-diag}, $A$
is diagnosable, \ssi, there is some $\Delta \in \setN$ \st $A$ is
$\Delta$-diagnosable. Thus:
\begin{eqnarray*}
  \label{eq-diagnos}
  \hskip-0em\text{$A$ \textbf{is not} diagnosable} &  \hskip-1.2em \iff   & 
  \hskip-1.2em \forall \Delta \in \setN, 
  \text{$A$ is not  $\Delta$-diagnosable.} 
\end{eqnarray*}
Moreover a trace based definition of $\Delta$-diagnosability can be
stated as: $A$ is $\Delta$-diagnosable \ssi
\begin{equation}
  \faulty_{\geq \Delta}^{\trace}(A) \cap
  \nonfaulty^{\trace}(A) = \emptyset \mathpunct. \label{eq-base}
\end{equation}
This gives a necessary and sufficient condition for
non-diagnosability and thus diagnosability:
\begin{eqnarray}
\label{eq-diagnos2}
\hskip0em\text{$A$ \textbf{is not} diagnosable} &  \hskip-1.2em \iff  
& \hskip-1.3em 
 \begin{cases}
   \forall \Delta \in \setN, \\
   \quad \exists \rho \in \nonfaulty(A) \\
   \quad \exists \rho' \in \faulty_{\geq \Delta}(A) \textit{ \st }  \\
   \quad\;\;\trace(\rho) =
   \trace(\rho') \mathpunct,
 \end{cases}
\end{eqnarray}
or in other words, there is no pair of runs $(\rho_1,\rho_2)$ with
$\rho_1 \in \faulty_{\geq \Delta}(A)$, $\rho_2 \in \nonfaulty(A)$ the
traces of which  are equal.

\section{Algorithms for Discrete Event Systems}\label{sec-des}
In this section we briefly review the main results about
diagnosability of discrete-event systems.  We consider here that the
DES is given by a FA $A=(Q,q_0,\Sigma_{\tauac,f},\rightarrow)$.

Moreover we assume that the automaton $A$ is such that every faulty
run of length $n$ can be extended to a run of length $n+1$; this
assumption simplifies the proofs (of some lemmas
in~\cite{cassez-fi-08}) and if $A$ does not satisfy it, it is easy to
add $\tauac$ loops to deadlock states of $A$ to ensure it holds. It
does not modify the observation made by the external observer and thus
does not modify the diagnosability status of $A$.

\subsection{Problem~\ref{prob-delta-diag}}
To check Problem~\ref{prob-delta-diag} we have to decide whe\-ther there is
a $(\Delta+1)$-faulty run $\rho_1$ and  a non-faulty run
$\rho_2$ that give the same observations when projected on $\Sigma$.
An easy way to do this is to build a finite automaton $\calB$ which
accepts exactly those runs, and check whether $\lang(\calB)$ is empty
or not.

Let $A_1=(Q \times
\{-1,0,\cdots,\Delta+1\},(q_0,-1),\Sigma_{\tauac},\rightarrow_1)$
be the automaton with $\rightarrow_1$ defined by:
\begin{itemize}
\item $(q,n) \xrightarrow{\ \lambda \ }_1 (q',n)$ if $q \xrightarrow{\
    \lambda \ } q'$ and $n=-1$ and $\lambda \in \Sigma \cup
  \{\tauac\}$;
\item $(q,n) \xrightarrow{\ \lambda \ }_1 (q',\min(n+1,\Delta+1))$ if
  $q \xrightarrow{\ \lambda \ } q'$ and $n \geq 0$ and $\lambda \in
  \Sigma \cup \{\tauac\}$;
\item $(q,n) \xrightarrow{\ \tauac \ }_1 (q',\min(n+1,\Delta+1)$ if
  $q \xrightarrow{\  f \ } q'$.
\end{itemize} 
Let $A_2=(Q,q_0,\Sigma_\tauac,\rightarrow_2)$ with: $q \xrightarrow{\
  \lambda \ }_2 q'$ if $q \xrightarrow{\ \lambda \ } q'$ and $\lambda
\in \Sigma \cup \{\tauac\}$.  Define $\calB=A_1 \times A_2$ with the
final states $F_\calB$ of $\calB$ given by:
$F_\calB=\{((\ell,\Delta+1),\ell') \ | \ (\ell,\ell') \in Q \times
Q\}$. We let $R_\calB=\emptyset$.  It is straightforward to see that:
\begin{theorem}
  $A$ is $\Delta$-diagnosable \ssi $\lang^*(\calB)=\emptyset$.
\end{theorem}

As language emptiness for $\calB$ amounts to reachability checking, it
can be done in linear time in the size of $\calB$.  Still strictly
speaking, the automaton $\calB$ has size $(\Delta+1) \cdot |A|^2$
which is exponential in the size of the inputs of the problem $A$ and
$\Delta$ because $\Delta$ is given in binary. Thus
Problem~\ref{prob-delta-diag} can be solved in EXPTIME.  As storing
$\Delta$ requires only polynomial space Problem~\ref{prob-delta-diag}
is in PSPACE.  Actually checking Problem~\ref{prob-delta-diag} can be
done in PTIME (see the end of this section).

\subsection{Problem~\ref{prob-diag}}
To check whether $A$ is diagnosable, we build a synchronized product
$A_1 \times A_2$, \st $A_1$ behaves exactly as $A$ but records in its
state whether a fault has occurred, and $A_2$ behaves like $A$ without
the faulty runs as before.  It is then as if $\Delta =\, 0$ in the
previous construction.
We let $\rightarrow_{1,2}$ be the transition relation of $A_1 \times
A_2$.  A faulty run of $A_1 \times A_2$ is a run for which $A_1$
reaches a faulty state of the form $(q,1)$. To decide whether $A$ is
diagnosable we build an extended version of $A_1 \times A_2$ which is
a B\"uchi automaton $\calB$ as follows: $\calB$ has a boolean variable
$z$ which records whether $A_1$ participated in the last transition
fired by $A_1 \times A_2$.  Assume we have a predicate\footnote{This
  is easy to define when building $A_1 \times A_2$.} $\aonemove(t)$
which is true when $A_1$ participates in a transition $t$ of the
product $A_1 \times A_2$.  A state of $\calB$ is a pair $(s,z)$ where
$s$ is a state of $A_1 \times A_2$.  $\calB$ is given by the tuple
$((Q \times \{0,1\} \times Q) \times
\{0,1\},((q_0,0),q_0,0),\Sigma_\tauac,\longrightarrow_{\calB},
\emptyset,R_{\calB})$ with:
\begin{itemize}
\item $(s,z) \xrightarrow{\ \sigma \ }_{\calB} (s',z')$ if $(i)$ there
  exists a transition $t: s \xrightarrow{\ \sigma \ }_{1,2} s'$ in
  $A_1 \times A_2$, and $(ii)$ $z'=1$ if $\aonemove(t)$ and $z'=0$
  otherwise;
\item $R_{\calB}=\{(((q,1),q'),1) \, | \, ((q,1),q') \in
A_1 \times A_2\}$.
\end{itemize}
$\calB$ accepts the language
$\lang(\calB)=\lang^\omega(\calB) \subseteq \Sigma^\omega$. 
Moreover this language satisfies a nice property:
\begin{theorem}[\cite{cassez-fi-08}] \label{thm-diagnosability}
  $A$ is diagnosable \ssi $\lang^\omega(\calB) = \emptyset$.
\end{theorem}
 
This theorem has for consequence that the diagnosability problem can
be checked in quadratic time: the automaton $\calB$ has size $4 \cdot
|A|^2$ \ie $O(|A|^2)$ and checking emptiness for B\"uchi automaton can
be done in linear time. Thus diagnosability can be checked in PTIME.
Polynomial algorithms for checking diagnosability
(Problem~\ref{prob-diag}) were already reported
in~\cite{Jiang-01,yoo-lafortune-tac-02}.  In these two papers, the
plant cannot have unobservable loops \ie loops that consist of
$\tauac$ actions.  Our algorithm does not have this limitation (we
even may have to add $\tauac$ loops to ensure that each faulty run can
be extended).  Note also that in~\cite{Jiang-01,yoo-lafortune-tac-02},
the product construction is symmetric in the sense that $A_2$ is a copy
of $A$ as well. Our $A_2$ does not contain the $f$ transitions, which
is a minor difference complexity-wise, but in practice this can be
useful to reduce the size of the product.

Moreover, reducing Problem~\ref{prob-diag} to emptiness checking of
B\"uchi automata is interesting in many respects:
\begin{itemize}
\item the proof (see~\cite{cassez-fi-08}) of
  Theorem~\ref{thm-diagnosability} is easy and short; algorithms
  for checking B\"uchi emptiness are well-known and
  correctness follows easily as well;
\item this also implies that standard tools from the
  \emph{model-checking/verification} community can be used to check
  for diagnosability. There are very efficient tools to check for
  B\"uchi emptiness (\eg
  \textsc{Spin}~\cite{spin-holzmann05}). Nu\-me\-rous algorithms, like
  \emph{on-the-fly} algorithms~\cite{couvreur-spin-05} have been designed
  to improve memory/time consumption (see~\cite{schwoon-tacas-05} for
  an overview).  Also when the DES is not diagnosable a
  counter-example is provided by these tools.  The input languages
  (like \textsc{Promela} for \textsc{Spin}) that can be used to
  specify the DES are more expressive than the specification languages
  of some dedicated tools\footnote{UMDES was the only publicly
    available tool which could be found by a Google search.}  like
  \textsc{DES\-UMA/UMDES}~\cite{lafortune-umdes} (notice that the
  comparison with DESUMA/UMDES concerns only the diagnosability
  algorithms; DES\-UMA/UMDES can perform a lot more than checking
  diagnosability).
\end{itemize}
    
From Theorem~\ref{thm-diagnosability}, one can also conclude that
diagnosability amounts to bounded diagnosability: indeed if $A$ is
diagnosable, there can be no accepting cycles of faulty states in
$\calB$; in this case there cannot be a faulty run of length more than
$2 \cdot |Q|^2$ in $\calB$. Thus Problem~\ref{prob-diag} reduces to a
particular instance of Problem~\ref{prob-delta-diag} which was already
stated in~\cite{yoo-lafortune-tac-02}:
\begin{theorem}[\cite{yoo-lafortune-tac-02}]\label{thm-reduc}
  $A$ is diagnosable if and only if  $A$ is $(2 \cdot |Q|^2)$-diagnosable. 
\end{theorem}
This appeals from some final remarks on the algorithms we should
choose to check diagnosability: for the particular case of
$\Delta=2\cdot|A|^2$, solving Problem~\ref{prob-delta-diag} (a
reachability problem) can be done in time $2\cdot|A|^2 \cdot |A|^2$
\ie $O(|A|^4)$ whereas solving directly Problem~\ref{prob-diag} as a
B\"uchi emptiness problem can be done in $O(|A|^2)$.  Thus the
extra-cost of using a reachability algorithm is still reasonable.

The B\"uchi-emptiness algorithm used to solve Problem~\ref{prob-diag}
can also be used to solve Problem~\ref{prob-delta-diag} for a given
$\Delta$ and automaton $A$ with set of states $Q$: if $\Delta \geq 2
\cdot |Q|^2$, then we check wether $A$ is diagnosable and this gives
the answer to Problem~\ref{prob-delta-diag}; otherwise, if $\Delta < 2
\cdot |Q|^2$, we check wether $A$ is $\Delta$-diagnosable but in
polynomial time. Hence Problem~\ref{prob-delta-diag} can be solved in
polynomial time $O(|A|^4)$.

Finally, solving Problem~\ref{prob-delay} can be done by a binary
search solving iteratively $\Delta$-diagnosability problems starting
with $\Delta=2\cdot|A|^2$. Thus Problem~\ref{prob-delay} can be solved
in $O(|A|^4)$.  Using a different approach, Problem~\ref{prob-delay}
was reported to be solvable in $O(|Q|^3)$ in~\cite{yoo-garcia-03}.

In the sequel we recall the algorithm for checking diagnosability for
TA and establish a counterpart of Theorem~\ref{thm-reduc} for TA.


\section{Algorithms for Timed Automata}\label{sec-ta}
We first recall how to check $\Delta$-diagnosability for TA which
first appeared in~\cite{tripakis-02}.

\subsection{Problem~\ref{prob-delta-diag}}
Let $t$ be a fresh clock not in $X$.
\noindent Let $A_1(\Delta)=((L \times \{0,1\} )\cup \{Bad\}
,(l_0,0),X \cup \{t\},\Sigma_\tauac,E_1,\inv_1)$ with:
\begin{itemize}
\item $((\ell,n),g,\lambda,r,(\ell',n)) \in E_1$ if
  $(\ell,g,\lambda,r,\ell') \in E$, $\lambda \in \Sigma \cup
  \{\tauac\}$;
\item $((\ell,0),g,\tauac,r \cup \{t\},(\ell',1)) \in E_1$ if
  $(\ell,g,f,r,\ell') \in E$;
\item $\inv_1((\ell,n))=\inv(\ell)$;
\item for $\ell \in L$, $((\ell,1),t \geq \Delta,\tauac,\emptyset,Bad)
  \in E_1$
\end{itemize} 
and $A_2=(L,l_0,X_2,\Sigma_\tauac,E_2,\inv_2)$ with: 
\begin{itemize}
\item $X_2 = \{x_2 \ | \ x \in X\}$ (clocks of $A$ are renamed);
\item $(\ell,g_2,\lambda,r_2,\ell') \in E_2$ if
  $(\ell,g,\lambda,r,\ell') \in E$, $\lambda \in \Sigma \cup
  \{\tauac\}$ with: $g_2$ is $g$ where the clocks $x$ in $X$ are
  replaced by their counterpart $x_2$; $r_2$ is $r$ with the same
  renaming;
\item $\inv_2(\ell)=\inv(\ell)$.
\end{itemize}

Consider $A_1(\Delta) \times A_2$.  A faulty state of $A_1(\Delta)
\times A_2$ is a state of the form $(((\ell,1),v),(\ell',v'))$ \ie
where the state of $A_1$ is faulty.  Let $\runs_{\geq
  \Delta}(A_1(\Delta) \times A_2)$ be the runs of $A_1(\Delta) \times
A_2$ \st a faulty state of $A_1$ is encountered and \st at least
$\Delta$ time units have elapsed after this state.  If this set is not
empty, there are two runs, one $\Delta$-faulty and one non-faulty
which give the same observation. Moreover, because $t$ is reset
exactly when the first fault occurs, we have $t \geq
\Delta$. Conversely, if a state of the form
$(((\ell,1),v),(\ell',v'))$ with $v(t) \geq \Delta$ is reachable, then
there are two runs, one $\Delta$-faulty and one non-faulty which give
the same observation. Location $Bad$ in $A_1$ is thus reachable
exactly if $A$ is not $\Delta$-diagnosable. Let $\calD$ be
$A_1(\Delta) \times A_2$ with the final set of locations
$F_\calD=\{Bad\}$ and $R_\calD=\emptyset$.
\begin{theorem}[\cite{tripakis-02}]
  $A$ is $\Delta$-diagnosable \ssi $\lang^*(\calD)=\emptyset$.
\end{theorem}
Checking reachability of a location for TA is
PSPACE-complete~\cite{AlurDill94}.  More precisely, it can be done in
linear time on the region graph.  The size of the region graph of
$\calD$ is $(2 \cdot |L|^2 + |L|) \cdot (2|X|+1)! \cdot 2^{2|X|+1}
\cdot K^{2|X|} \cdot \Delta$ where $K$ is the maximal constant
appearing in $A$. Hence:
%
\begin{corollary}
  Problem~\ref{prob-delta-diag} can be solved in
  PSPACE for TA.
\end{corollary}

\subsection{Problem~\ref{prob-diag}}

\begin{figure}[hbtp]
\centering
  \begin{tikzpicture}[thick,node distance=1cm and 4cm,bend angle=20]%
    \small
    \node[state,initial] (q_0) [label=-93:{$[x \leq 1]$}] {$0$}; 
    \node[state] (q_1) [right=of q_0,label=-87:{$[x \leq 1]$}] {$1$};
    \path[->] (q_0) edge [bend left] node[pos=0.5] {$x=1$; $\tauac$; $x:=0$} (q_1) 
              (q_1) edge [bend left] node[pos=0.5] {$x=1$; $\tauac$; $x:=0$} (q_0);
  \end{tikzpicture}
\caption{Timed Automaton $\dive(x)$}
 \label{fig-dive}  
\end{figure}
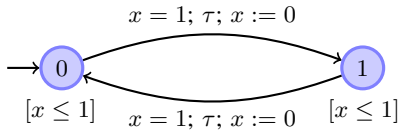

As for the untimed case, we build an automaton $\calD$, which is
special version of $A_1(\Delta) \times A_2$.  Assume $A_1$ is defined
as before omitting the clock $t$ and the location $Bad$.
In the timed case, we have to take care of the following real-time
related problems~\cite{tripakis-02}:
%
\begin{itemize}
\item some runs of $A_2$ might prevent time from elapsing from a given
  point in time. In this case, equation~(\ref{eq-diagnos}) cannot be
  satisfied but this is for an artificial reason: for $\Delta$ large
  enough, there will be no $\Delta$ faulty run in $A_1 \times A_2$
  because $A_2$ will block the time. In this case we can claim that
  $A$ is diagnosable but it is not realistic;
\item a more tricky thing may happen: $A_1$ could produce a Zeno
  run\footnote{A Zeno run is a run with infinitely many discrete steps
    the duration of which is bounded.}  after a fault occurred. This
  could happen by firing infinitely many $\tauac$ transitions in a
  bounded amount of time.  If we declare that $A$ is not diagnosable
  but the only witness run is a Zeno run, it does not have any
  physical meaning. Thus to declare that $A$ is not diagnosable, we
  should find a non-Zeno witness which is realizable, and for which
  time diverges.
\end{itemize}
To cope with the previous dense-time related problems we have to
ensure that the two following conditions are met:
\begin{description}
\item[$C_1$:] $A_2$ is \emph{timelock-free} \ie $A_2$ cannot prevent
  time from elapsing; this implies that every finite non-faulty run of
  $A_2$ can be extended in a time divergent run. We can assume that
  $A_2$ satisfies this property or check it on $A_2$ before checking
  diagnosability;
\item[$C_2$:] for $A$ to be non-diagnosable, we must find an infinite
  run in $A_1 \times A_2$ for which time diverges.
\end{description}
$C_2$ can be enforced by adding a third timed automaton $\dive(x)$ and
synchronizing it with $A_1 \times A_2$. Let $x$ be a fresh clock not
in $X$. Let $\dive(x)=(\{0,1\},0,\{x\},E,\inv)$ be the TA given in
Fig.~\ref{fig-dive}.
If we use $F=\emptyset$ and $R=\{1\}$ for $\dive(x)$,
any accepted run is time divergent.
Let $\calD=(A_1 \times A_2) \times \dive(x)$ with 
$F_{\calD}=\emptyset$ and $R_{\calD}$ is the set of states where $A_1$
is in a faulty state and $\dive(x)$ is location~$1$.
%
%
The following theorem is the TA counterpart of
Theorem~\ref{thm-diagnosability}:
\begin{theorem}[\cite{tripakis-02}]\label{thm-diag-tempo}
  $A$ is diagnosable \ssi $\lang^\omega(\calD)= \emptyset$.
\end{theorem}
Deciding whether $\lang^\omega(A) \neq \emptyset$ for TA
is PSPACE-complete~\cite{AlurDill94}. Thus deciding diagnosability is
in PSPACE.

The reachability problem for TA can be reduced to a diagnosability
problem~\cite{tripakis-02}.  Let $A$ be a TA on alphabet $\Sigma$ and
$\textit{End}$ a particular location of $A$. We want to check whether
$\textit{End}$ is reachable in $A$.  It suffices to build $A'$ on the
alphabet $\Sigma_{\tauac,f}$ by adding to $A$ the following
transitions: $(\textit{End},\true,\lambda,\emptyset,\textit{End})$ for
$\lambda \in \{\tauac,f\}$.  Then: $A'$ is not diagnosable \ssi
$\textit{End}$ is reachable in $A$. It follows that:
\begin{theorem}[\cite{tripakis-02}]
  Problem~\ref{prob-diag} is PSPACE-complete for TA.
\end{theorem}
We can draw another conclusion from the previous theorem: if a TA $A$
is diagnosable, there cannot be any cycle with faulty states in the
region graph of $A_1 \times A_2 \times \dive(x)$.  Indeed, otherwise,
by Theorem~\ref{thm-alur}, there would be a non-Zeno word in $A_1
\times A_2 \times \dive(x)$ itself\footnote{Note that this is true
  because we add the automaton $\dive(x)$. Otherwise an infinite run
  in the region graph of a TA does not imply a time divergent run in
  the TA $A$ itself.}. Let $\alpha(A)$ denote the size of the region
graph $\rg(A_1 \times A_2 \times \dive(x))$. If $A$ is diagnosable,
then ($P_1$): a faulty state in $\rg(A_1 \times A_2 \times \dive(x))$
can be followed by at most $\alpha(A)$ (faulty) states. Notice that a
faulty state cannot be followed by a state $(s,r)$ where $r$ is an
unbounded region of $A$, as this would give rise to a non-Zeno word in
$A_1 \times A_2 \times \dive(x)$. Hence ($P_2$): all the regions
following a faulty state in $\rg(A_1 \times A_2 \times \dive(x))$ are
bounded.  As the amount of time which can elapse within a region is
less than $1$ time unit\footnote{We assume the constants are
  integers.}, this implies that the duration of the longest faulty run
in $A_1 \times A_2 \times \dive(x)$ is less than $\alpha(A)$. Actually
as every other region is a \emph{singular region}\footnote{A singular
  region is a region in which time elapsing is not possible \eg
  defined by $x=0 \wedge y \geq 1$. }, it must be less than
$(\alpha(A)/2) +1$.  Thus we obtain the following result:
\begin{theorem}\label{thm-reduc-ta}
  $A$ is diagnosable if and only if $A$ is $(\alpha(A)/2)+1$-diagnosable.
\end{theorem}
As diagnosability can be reduced to $\Delta$-diagnosability for TA:
\begin{corollary}\label{cor-reduc-ta}
  Problem~\ref{prob-delta-diag} is PSPACE-complete for TA.
\end{corollary}
Problem~\ref{prob-delay} can be solved by a binary search and is also
in PSPACE for TA.
Although Problem~\ref{prob-delta-diag} and Problem~\ref{prob-diag} are
PSPACE-com\-plete for timed automata, the price to pay to solve
Problem~\ref{prob-diag} as a reachability problem is much higher than
solving it as a B\"uchi emptiness problem: indeed the size of the
region graph of $A_1(\alpha(A)) \times A_2$ is the square of the size
of the region graph of $A_1 \times A_2 \times \dive(x)$ which is
already exponential in the size of $A$.  Time-wise this means a blow
up from $2^n$ to $2^{n^2}$ which is not negligible as in the discrete
case.

\section{Conclusion}\label{sec-conclu}
\addtolength{\textheight}{-5.9cm}
\noindent The main conclusions we can draw from the previous
presentation are two-fold.

From a theoretical viewpoint, it shows that the fault diagnosis
algorithms for DES and for TA are essentially the same: in both cases,
diagnosability can be reduced to B\"uchi emptiness; and also to
bounded diagnosability. The interesting point is that the complexity
of the algorithms are the same for DES and TA except that for timed
automata, the complexity measure is space (Table~\ref{tab-summary}).
 \newcommand{\vtab}[1]{
  \begin{tabular}[c]{c}
    #1
  \end{tabular}
}
\begin{table}[tbhp]
  \centering
  \caption{Summary of the Results}
  \label{tab-summary}
  \begin{tabular}[t]{||l|c|c|c||}\hline\hline
    &  $\Delta$-Diagnosability   & \multicolumn{2}{c||}{Diagnosability} \\\cline{3-4}
    & $\textit{Reach Algorithm}$  & $\textit{B\"uchi Algorithm}$ & $\textit{Reach Algorithm}$ \\\hline\hline
    DES & \vtab{PTIME \\ $O(|A|^4)$} & \vtab{PTIME \\ $O(|A|^2)$} & \vtab{PTIME \\ $O(|A|^4)$}   \\\hline
    TA & PSPACE-C. & \vtab{PSPACE-C. \\ $O(|A|^2)$} &  \vtab{PSPACE-C. \\ $O(|A|^4)$}\\\hline\hline
  \end{tabular}
\end{table}

From a practical viewpoint, it clearly shows that the model-checking
algorithms and tools developed in the model-checking/verification
community can be used to solve the diagnosability problems; these
tools usually have a very expressive specification language (\eg
Promela/Spin~\cite{holzmann05}, UPPAAL~\cite{uppaal-tutorial04} or
KRONOS~\cite{kronos}) and very efficient data
structures/implementations (\eg~\cite{schwoon-tacas-05}
or~\cite{bbdlpw-ftrtft02}).

We can also use the results in Table~\ref{tab-summary} to guide our
choice of algorithms for checking diagnosability.  Let
$\textit{Reach}$ denote the reachability algorithm for checking
$\Delta$-diagnosability and $\textit{Buchi}$ denote the B\"uchi
emptiness algorithm for checking diagnosability:
\begin{itemize}
\item time-wise, solving the diagnosability problem for a finite
  automaton using $\textit{Reach}$ is a bit more expensive than using
  $\textit{Buchi}$, but the difference is not drastic;
\item for a timed automaton $A$ it is totally different: space-wise
  the amount of space required by $\textit{Reach}$ is the square of
  the amount of space required by $\textit{Buchi}$. Time-wise this
  means a worst case blow up from $2^{|A|^2}$ to $2^{|A|^4}$.  It is
  thus clear that one should use the B\"uchi emptiness algorithm in
  this case. Checking B\"uchi emptiness for TA is efficiently
  implemented in a version of KRONOS
  (Profounder)~\cite{tripakis-acm-09} and in
  UPPAAL-TiGA~\cite{uppaal-tiga}, the game version of
  UPPAAL~\cite{david-cav-09}.
\end{itemize}
The previous results show that model-checking tools (both for finite
and timed automata) are suitable to solve the diagnosis problems, and
provide expressive specification languages and efficient algorithms
and tools.

\bibliographystyle{IEEEtran}
\bibliography{diagnosis}

\end{document}